\documentclass[%
 reprint,
nofootinbib,
 amsmath,amssymb,
 aps,
]{revtex4-2}

\usepackage{graphicx}% Include figure files
\usepackage{dcolumn}% Align table columns on decimal point
\usepackage{bm}% bold math
\usepackage{bbold}
\usepackage{amsthm}
\usepackage{mathtools}
\usepackage{physics}
\usepackage{xcolor}
\usepackage{graphicx}
\usepackage{adjustbox}
\usepackage{placeins}
\usepackage[T1]{fontenc}
\usepackage{lipsum}
\usepackage{csquotes}
\usepackage{tikz}
\usepackage{float}
\usetikzlibrary{quantikz}
\usepackage{amsmath}
\usepackage{amssymb}
\usepackage{esvect}
\usepackage{tikz}
\usepackage{caption}
\usepackage{subcaption}
\usetikzlibrary{shapes.geometric, arrows}
\usetikzlibrary{shapes,arrows}
\tikzset{
    myarrow/.style={
        draw,
        fill=red,
        single arrow,
        minimum height=2.5ex,
        line width=1pt,
        single arrow head extend=0.1ex
    }
}

\tikzstyle{startstop} = [rectangle, rounded corners, 
minimum width=0.1cm, 
minimum height=0.5cm,
text width = 4cm,
text centered, 
draw=black, 
fill=red!30]

\tikzstyle{io} = [trapezium, 
trapezium stretches=true, % A later addition
trapezium left angle=35, 
trapezium right angle=35, 
minimum width=3cm, 
minimum height=1cm, text centered, 
draw=black, fill=blue!30]

\tikzstyle{process} = [rectangle, 
minimum width=1cm, 
minimum height=0.5cm, 
text centered, 
text width=2.5cm, 
draw=black, 
fill=orange!30]

\tikzstyle{decision} = [diamond, 
minimum width=0.5cm, 
minimum height=0.5cm, 
text centered, 
draw=black, 
fill=green!30]
\tikzstyle{arrow} = [thick,->,>=stealth]

\begin{document}

\preprint{APS/123-QED}

% \title{Manuscript Title:\\with Forced Linebreak}% Force line breaks with \\
\title{Randomized term grouping over physical law \\ on digital quantum simulation}
% \thanks{A footnote to the article title}%

\author{Songqinghao Yang}
 \altaffiliation[Also at ]{Cavendish Laboratory, University of Cambridge}%Lines break automatically or can be forced with \\
\affiliation{%
 Yusuf Hamied Department of Chemistry\\
 % This line break forced with \textbackslash\textbackslash
 Univeristy of Cambridge, United Kingdom\\
 E-mail: sqhy@cam.ac.uk
}%

\date{\today}% It is always \today, today,
             %  but any date may be explicitly specified

\begin{abstract}
We introduce a randomized algorithm based on qDrift to compute Hamiltonian dynamics on digital quantum computers. We frame it as physDrift because conservation laws in physics are obeyed during the evolution of arbitrary quantum states. Empirically, we achieved better spectral error reduction with the hydrogen chain model compared to previous protocols. Noisy models are investigated as well, and we characterized them in the circuit with different schemes, i.e. attenuation of the measured expectation value is fixed by keeping the circuit depth the same and depolarising error is simulated with randomly applied Pauli gates. This makes our proposal particularly feasible for implementing and testing on present-day noisy hardware.

\end{abstract}

%\keywords{Suggested keywords}%Use showkeys class option if keyword
                              %display desired
\maketitle

%\tableofcontents

% \section{\label{sec:level1}First-level heading:\protect\\ The line
% break was forced \lowercase{via} \textbackslash\textbackslash}
\section{\label{sec:level1} Introduction}
One of the advantages of quantum computing over classical computation is to simulate complex quantum systems. This idea was originally proposed famously by Richard Feynman\cite{feyn82}:
\begin{quotation}
    "Let the computer itself be built of quantum mechanical elements which obey quantum mechanical laws."
\end{quotation}
In the past few decades, various algorithms have been put forward. Seth Llodys\cite{lloyd1996universal} first formulated the problem for local Hamiltonian simulation where k-local means, for $H = \sum_j^{L} H_j$, each term acts on at most k qubits rather than all n qubits\footnote{In this case the problem is reduced to polynomial space: $L = \text{poly}(n)$ since $L \leq \sum_{j=1}^{k} {n \choose j} \leq c {n \choose j} \leq \frac{n^{n-c}}{(c-1)!}$}: given any Hamiltonian evolution unitary $U = e^{-iHt}$, we can find a set of quantum gates that could approximate the evolution:
\begin{equation}
    U \approx V = V_1 V_2 \dots V_N.
\end{equation}
He then explicitly used a compilation method called product formula or trotterization to find out how the error $\epsilon(U) = \Vert e^{-iHt} - V \Vert$\footnote{Here spectral norm is taken as the measurement metric between different unitaries. We also consider real-time evolution(RTE) throughout the paper.}, scales with system parameters: Consider many-body Hamiltonian as $H = \sum_{k=1}^{L} h_k H_k$, where $H_k$ is one sub-term (usually a Pauli string) in the Hamiltonian and $h_k$ is the associated coefficient. The first order trotterization (or lie trotter) is defined as:
\begin{equation}
    S_1(t) = \prod_{k=1}^L \exp ( -i h_k H_k t).
\end{equation}
The scaling follows:
\begin{equation}
    N_t = \mathcal{O}(\frac{(t L \Lambda )^2}{\epsilon}),
\end{equation}
where $\Lambda := \max_k \Vert h_k \Vert$.
So, to make a better approximation, we need to increase the time steps $N_t$:
\begin{equation}
    e^{-i\sum_{k=1}^{L} h_k H_k} \approx S_1(\frac{t}{N})^N = S_1(\delta t) S_1(\delta t) \dots S_1(\delta t).
\end{equation}
Built upon that, higher order\footnote{Suzuki product formulae are symmetric and therefore only defined for even orders.} formula is derived for better scaling\cite{hatano2005finding} in terms of Trotter errors but comes with a trade-off between the number of gates required and precision per Trotter step, and we can demonstrate that the balance is optimized around $2^{nd}$ and $4^{th}$ order in Appendix\ref{apB}:
\begin{eqnarray}
    S_2(t) = \prod_{k=1}^L \exp ( -i h_k H_k \frac{t}{2})\prod_{k=L}^1 \exp ( -i h_k H_k \frac{t}{2}),
\end{eqnarray}
\begin{eqnarray}
    S_{2k}(t) = S_{2k - 2}(p_kt)^2S_{2k - 2}\left( (1-4p_k)t\right)S_{2k - 2}(p_kt)^2,
\end{eqnarray}
where $p_k = 1/(4 - 4^{1/(2k - 1)})$ and they scale as:
\begin{equation}
    N = \mathcal{O}(\frac{(t L \Lambda )^{1+\frac{1}{2k}}}{\epsilon^{1/2k}}).
\end{equation}
Better accuracy comes from the fact that some of the error terms in non-commutativity will be cancelled with a recursive formula in both forward and backward directions. More advanced techniques like classical optimization\cite{mc2023classically}, a linear combination of unitaries(LCU)\cite{childs2012hamiltonian}, quantum signal processing\cite{low2017optimal} and truncated Taylor series\cite{berry2015simulating} have demonstrated improvements in different aspects afterwards. For example, with the truncation order for LCU, the number of gates needed per step is approximately: 
\begin{equation}
    \mathcal{K} = \mathcal{O}(\frac{log(\frac{1}{\epsilon})}{loglog(\frac{1}{\epsilon})}),
\end{equation}
while Trotterization has the exponential dependency, i.e. fourth-order $\mathcal{O}(\frac{1}{\epsilon^{1/4}})$.  Not only does the Taylor series give logarithmically better scaling, but it also comes with a better empirical result\cite{childs2021theory}(tightening the bound between theoretical prediction and empirical results has been heavily discussed\cite{childs2019nearly}). We categorized them in FIG.1 above. 

However, product formulas remain the most popular, particularly for near-term hardware simulation\cite{barends2015digital,brown2006limitations}. This is because:
\begin{enumerate}
    \item Norm of the wavefunction is preserved: errors in expectation values are expected to oscillate, vs in non-unitary evolution errors accumulate, e.g. $$\| 1-iHt \| \approx 1 + \|H\|t > 1$$ with LCU.
    \item It has simple implmentation with no overhead (unitary operation can be implemented deterministically, without ancilla qubits).
\end{enumerate}
\begin{figure}
\includegraphics[scale=0.27]{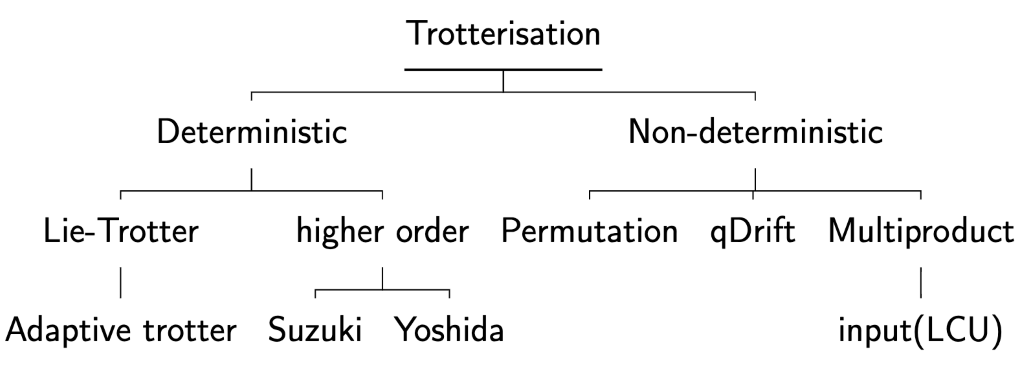}% Here is how to import EPS art
\caption{Summarized state-of-art method in digital quantum simulation (DQS). We mention alternative approaches briefly. Adaptive trotter\cite{zhang2023low} framework was recently proposed with better empirical error bound; however, this method involves feed-forwarding circuits and is unsuitable for near-term application. Yoshida\cite{yoshida1990construction} used a leapfrog scheme for higher orders, and it has been improved since\cite{barthel2020optimized}. With a recent proposal on specific conditions\cite {low2019well} to reduce the exponential parameter dependency in Multiproduct formula\cite{childs2012hamiltonian}, the approach of parallelization of product formula has become more practical.}
\end{figure}
However, despite achieving tighter error bound\cite{childs2021theory} for trotterization and progress in tuning the parameter, e.g. $p_k$\cite{jones2019optimising} to make the simulation more accurate, problems remain with the intrinsic property of the algorithm: the exponential dependency of the recursive form $\mathcal{O}(5^k)$ and the inclusion of the number of Hamiltonian terms L. The second issue is particularly prominent in extensive systems like quantum chemistry, where complex electronic structures involve a lot of Pauli strings after some transformation.

Consequently, several protocols focused on non-deterministic construction. In contrast to deterministic algorithms that produce unitary quantum circuits, randomized algorithms either randomly permute the ordering of unitary gates during simulation\cite{childs2019faster}, which we will refer to 'random permutation', or utilize mixing lemma\cite{campbell2017shorter, hastings2016turning} on quantum channels (qDrift\cite{campbell2019random}) to make better asymptotic scaling. In this paper, we will concentrate on these two significant approaches and demonstrate our algorithm's advantage by theoretically and empirically combining physical laws. Specifically, we show that simulating the evolution while preserving particle number makes the state-vector more accurate, which we refer to as physDrift. We will show subsequently that this could be achieved by term grouping. Deterministic methods are also included for benchmarking. We demonstrate that physDrift has smaller errors and significantly reduces the leakage to unphysical systems in both short and long time. That is to say, the conservation law is obeyed during evolution. Finally, we must consider errors when implementing the circuits on real hardware. So, the depolarising error is added, and the simulation is again benchmarked for comparison.

In the rest of the paper, we will first provide some background on the model we use in section \ref{sec2}. Then, we briefly review qDrift and random permutation in section \ref{sec3}. In section \ref{sec4}, the protocol for physDrift is laid out, and we evaluate the theoretical and empirical results with other algorithms. In the second part of the section, the symmetric protection technique is introduced for deterministic methods, and we benchmark all algorithms on real chemical systems, e.g. hydrogen chains. We summarise in section \ref{sec5} and comment on possible future improvements.

\section{Background}\label{sec2}

Hamiltonian evolution operator is too complex to be directly implemented on quantum circuits. This section explains the method used to decompose and transform the unitary $e^{-iHt}$ and then integrate it with near-term hardware.

\subsection{Chemistry model}

We focus on electronic structure problems, which could be initialized with the electronic model (Molecular Hamiltonian or Fermi-Hubbard model in physics):
\begin{equation}
    H=\sum_{pq}h_{pq}a^\dagger_{p}a_{q}+\frac{1}{2}\sum_{pqrs}h_{pqrs}a^\dagger_{p}a^\dagger_{q}a_{r}a_{s}+h_{nuc},
\end{equation}
where $h_{nuc}$ is the nuclear energy, we ignore it in the Born-Oppenheimer approximation, making it a constant. $a_p, a_p^\dag$ are fermionic operators satisfying:
\begin{equation}
    a^\dagger_1 \ket{0}_1 = \ket{1}_1,\quad a^\dagger_1 \ket{1}_1 = 0,\quad a_1 \ket{0}_1 = 0,\quad a_1 \ket{1}_1 = \ket{0}_1.
\end{equation}
Moreover, they anti-commute with operators with different labels.
The coefficients are defined as:
\begin{eqnarray}
    h_{pq} = \int_{-\infty}^\infty \psi^*_p(x_1) \left(-\frac{\nabla^2}{2} +V(x_1)\right)  \psi_q(x_1)\mathrm{d}^3x_1.
\end{eqnarray}
\begin{widetext}    
\begin{eqnarray}
h_{pqrs} = \int_{-\infty}^\infty \int_{-\infty}^\infty\psi_p^*(x_1)\psi_q^*(x_2) \left(\frac{1}{|x_1-x_2|} \right)\psi_r(x_2)\psi_s(x_1)\mathrm{d}^3x_1\mathrm{d}^3x_2,
\end{eqnarray}
\end{widetext}
where $V(x)$ is the mean-field potential. $h_{pq}$ generally represents the integral containing kinetic energy + electron-nucleus attraction while $h_{pqrs}$ is the two-body interaction term.
Together with the creation and annihilation operator, we can classify terms in Molecular Hamiltonian as in TABLE.\ref{table:ta1}.

In particular, we have integrated $h_{pqpq}$ into $h_{pqqp}$. $h_{pqpq}$ is an exchange coupling - it is only non-zero when the two electrons have the same spin so that they have an antisymmetric spatial state, which reduces the coulomb repulsion. So, this term tends to induce ferromagnetic coupling between neighbouring spins. As we make the separation between sites bigger, the $h_{pqpq}$ term gets exponentially smaller; we can consider only the $h_{pqqp}$ term. Correlated excitation between electrons means that the interaction depends on the location of the electrons, i.e. if electron 1 in r jumps to p, electron 2 in q stays in q. This can be seen directly from the expansion of the operator using fermionic anti-commutation relations:
\begin{equation}
    a^\dagger_{p}a^\dagger_{q}a_{q}a_{r} = (a^\dagger_{p}a_{r} + a^\dagger_{r}a_{p})(a^\dagger_{q}a_{q}).
\end{equation}
\begin{table}[t] 
\caption{
The five classes of sub-Hamiltonian with second quantized operators. In general, there are few terms in each class because of symmetric reduction.}
\label{table:ta1}
\begin{ruledtabular}
\begin{tabular}{cc}
Physical meaning&Operator\\
\hline
electron number counting&$h_{pq}a^\dagger_{p}a_{p}$\\
electron excitation\footnote{Note the model expresses Hamiltonians not in localized basis (i.e. orthonormal atomic orbitals), instead it is expressed using molecular orbitals (i.e. linear combinations of atomic orbitals, where the expansion coefficients are found using Hartree Fock). So, we cannot use terms like 'hopping' between sites; we can only use excitation between molecular orbits.}&$h_{pq}(a^\dagger_{p}a_{q} + a^\dagger_{q}a_{p}$)\\
Coulomb repulsion & $h_{pqqp}a^\dagger_{p}a^\dagger_{q}a_{q}a_{p}$\\
correlated excitation & $h_{pqqr}a^\dagger_{p}a^\dagger_{q}a_{q}a_{r}$\\
Scatter & $h_{pqrs}a^\dagger_{p}a^\dagger_{q}a_{r}a_{s}$
\end{tabular}
\end{ruledtabular}
\end{table}
\subsection{Transformation}

Using the Jordan-Wigner transformation\cite{jordan1993paulische}, we can map the fermionic operator to the space spanned by Pauli operators:
\begin{equation}
    a^\dagger_j = \begin{bmatrix} 0 & 0 \\ 1 & 0 \end{bmatrix}=\frac{X_j - iY_j}{2},
\end{equation}
\begin{equation}
    a_j= \begin{bmatrix} 0 & 1 \\ 0 & 0 \end{bmatrix}=\frac{X_j + iY_j}{2}.
\end{equation}
To give a concrete example, consider the excitation term:
\begin{equation}
    h_{pq}(a_p^\dagger a_q + a^\dagger_q a_p) = \frac{h_{pq}}{2}\left(\prod_{j=q+1}^{p-1} Z_j \right)\left( X_pX_q + Y_pY_q\right).
\end{equation}
Other mappings like Bravyi Kitaev transformation\cite{seeley2012bravyi} and parity transformtion\cite{bravyi2017tapering} are also available. However, we will stick to the most straightforward approach and leave improvements for future work.

\subsection{Pauli gadgets}

With Hamiltonian as Pauli strings in hand, we consider the circuit in FIG.3, which is the controlled rotational gate in Z.
\newcommand{\RTEcirc}[3]{
\begin{figure}[h!]
\centering
\scalebox{#1}{
\begin{quantikz}[column sep=#2cm, row sep={#3cm,between origins}]
    %\lstick[wires=4]{$\ket{\psi}$}
     & \qw & \qw & \ctrl{1} & \qw & \qw & \qw & \ctrl{1} & \qw & \qw & \qw \\
     & \qw & \qw & \targ{} & \ctrl{1} & \qw & \ctrl{1} & \targ{} & \qw & \qw & \qw \\
    &\qw & \qw & \qw & \targ{} & \gate{R_z(2t)} & \targ{} & \qw & \qw & \qw & \qw
\end{quantikz}
}
\caption{Quantum circuit realisation of $U(t) = e^{-it ZZZ}$.}.
\end{figure}
}
\RTEcirc{1}{0.25}{0.9}
where the CNOT gates take the role of parity checkers connecting all qubits together. The factor of 2 in the rotational gate comes from the fact that the Pauli group, which belongs to SU(2), is actually a double cover of SO(3).

We need to use the relation $HZH = X$ and $Y = (SH)Z(SH)^\dag$, where H is the Hadamard gate, and S is the phase gate, to compute universal Pauli strings as in FIG.3.
\begin{figure}
\includegraphics[scale=0.5]{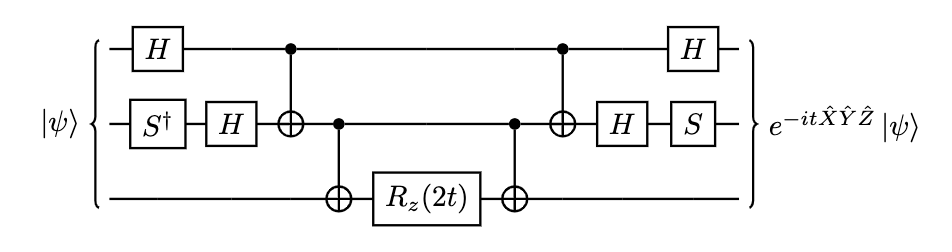}
\caption{Quantum circuit realisation of  $U(t) = e^{-it XYZ}$. }
\end{figure}
To simulate all Pauli strings, we need to concatenate our circuits using Theorem\ref{thm1}
\newtheorem{theorem}{Theorem}
\begin{theorem}\label{thm1}
For any two sub-terms $H_i$, $H_j$ in Hamiltonian $H = \sum_{k}^{L} h_k H_k$, if they satisfy,
\begin{equation}
    [H_i, H_j] = 0,
\end{equation}
then we have:
\begin{equation}
    e^{-i(H_i+H_j)t} = e^{-iH_it}e^{-iH_jt}.
\end{equation}
\end{theorem}
So, if the Pauli strings commute with one another, directly connecting Pauli gadgets will give us the exact evolution operator. We also get the intuition that with more commutative terms, the error goes down. Bound has been proved through both commutator\cite{childs2021theory} and anti-commutator\cite{zhao2021exploiting} relations.

\section{Randomized protocols}\label{sec3}

In this section, we begin by reviewing the density matrix: A classical distribution $p_i$ of quantum states gives a mixed state and must be described using a density matrix,
    $$\rho = \sum_{i=1}^{m} p_i \underbrace{\ket{i}}_{\text{pure state}} \bra{i}.$$
If $m>1$, $\rho$ is a mixed density matrix.
Furthermore, we remind you that for any $\rho$, the following properties are obeyed:
\begin{itemize}
    \item $Tr(\rho) = 1$
    \item $\rho > 0$
    \item Hermitian
\end{itemize}
We also clarify some of the notations. All capital curly letters refer to the quantum channel, which correspondingly evolves density matrix by unitary operators, e.g. $\mathcal{U} = e^{iHt}\rho e^{-iHt}$. In the remaining paper, we use $\mathcal{U}$ as the actual evolution channel and $\mathcal{V}, \mathcal{E}$ as approximations. We will sometimes utilize liouvillian representation for quantum channel:
\begin{equation}
    e^{iHt} \rho e^{-iHt} = e^{t\mathcal{L}} = \sum_{k=0}^\infty \frac{t^k \mathcal{L}^k (\rho)}{k!},
\end{equation}
where $\mathcal{L} (\rho) =i (H\rho - \rho H) $, similarly. Moreover, we have, similarly as above, the composition of Hamiltonians with sub-terms:
\begin{eqnarray}
    \mathcal{L} (\rho) = \sum_{j} h_j \mathcal{L}_j,
\end{eqnarray}
\begin{eqnarray}
    \mathcal{L}_ j(\rho)=i(H_j\rho - \rho H_j).
\end{eqnarray}
A quantum channel describes how a particular state or, more usefully, a mixture of quantum states evolves into another set of quantum states. The most general `channel' is a CPTP(completely positive, trace preserving) map, i.e. something that maps a density matrix onto another density matrix:
\begin{equation}
    \rho \rightarrow \sum_i M_i^\dag \rho M_i, \text{ where $\sum_i M_i^\dag M_i = 1$},
\end{equation}
where $M_i$ is called Kraus operator\cite{Kraus}. Thus, classically sampling a series of exponentials defines a valid CPTP map with $M_i = \sqrt{p_i} U_i$:
\begin{equation}
    \rho \rightarrow \sum_i p_i U_i^\dag \rho U_i.
\end{equation}
If we start in a pure state, essentially, we can generate the following mixed state:
\begin{equation}
    \ket{\psi}\bra{\psi} \rightarrow \sum_i p_i U_i^\dag \ket{\psi}\bra{\psi} U_i = \sum_i p_i \ket{\Psi_i}\bra{\Psi_i}.
\end{equation}

\subsection{Metric}

Now, we introduce some useful results to help the analysis of randomized algorithms. We use the diamond norm to calculate the difference between the simulated state-vector and the actual one:
\begin{equation}
    d_\diamondsuit (\mathcal{E},\mathcal{U}) =\frac{1}{2} \Vert \mathcal{E} -\mathcal{U} \Vert_\diamondsuit = \sup_{\Vert \rho \Vert_1 =1} \frac{1}{2} \Vert ((\mathcal{E}-\mathcal{U}) \otimes \mathbb{I})(\rho )\Vert_1,
\end{equation}
where $\mathbb {I} $ is the identity channel which has the same size as $\mathcal{E}$ and $\Vert \cdot \Vert$ is Schatten-$1$ norm or trace norm. Trace norm can be considered as the Euclidean distance between two quantum states:
\begin{eqnarray}
    \Vert M \Vert = \operatorname {tr} (\sqrt{MM^\dag}),
\end{eqnarray}
\begin{eqnarray}
    M = \ket{\psi}\bra{\psi} -\ket{\phi}\bra{\phi}.
\end{eqnarray}
Following the above, we can show that given an operator M,
\begin{equation}
    | \operatorname {tr} (M\mathcal{E}) - \operatorname {tr} (M\mathcal{U}) | \leq 2 \Vert M \Vert d_\diamondsuit (\mathcal{E},\mathcal{U}).
\end{equation}
It is important to note that because the diamond norm evaluates the maximum possibility that the two channels can be distinguished in \textit{all quantum states}(meaning that the smaller the value, the closer the simulation is to reality), this is the worst scenario bound for measuring the error because effectively it is evaluating the error over all state vector space. Although other spectral error metrics have been assessed intensively\cite{yi2022spectral}, we still adopt the diamond norm as it is the most commonly used one in the literature. As each quantum gate represents a unitary operator, to relate it to the quantum channel, we employ a mixing lemma:

\newtheorem{lemma}{Lemma}
\begin{lemma}\label{lem1}
Let $\mathcal{U}=U\rho U^\dag$ and $\mathcal{V}_j=V_j\rho V_j^\dag$ be unitary channels, and let ${p_j}$ be the probability distribution for the randomized protocol, suppose that:\\
$$\Vert \sum_{j}(p_jV_j) - U\Vert \leq a, $$
Then the mixed channel $\mathcal{V}:=\sum_{j}(p_j\mathcal{V_j})$ satisfy
$$ \Vert \mathcal{V} - \mathcal{U} \Vert \leq 2a. $$
\end{lemma}
Note that this is an improved version of the original lemma, and the proof is in Appendix\ref{apA} as well as [Lemma 3.4]\cite{chen2021concentration}.

\subsection{qDrift}

With Lemma\ref{lem1}, we first choose an important sampling distribution ${p_j}$. Here we consider the original construction, i.e. $p_j = \frac{|h_j|}{\lambda}$ where $|\cdot|$ means absolute value and $\lambda = \sum_{j=1}^{L} |h_j|$ . More advanced important sampling techniques could be used to reduce the cost of the circuit\cite{kiss2023importance, tokdar2010importance} in Appendix\ref{apC}. Then for each sample step\footnote{qDrift does not have the idea of the time step. Instead, a number of Pauli strings sampled specifies the interval of evolution.} we sample from the pool ${H_j}$ and implement the Pauli string with modified coefficients $\lambda t / N$ with N being the total sample number. Effectively, with probability $\lambda^N \prod_{j=1}^N |h_{j_k}|$, we have constructed a quantum channel:
\begin{eqnarray}
    \mathcal{V} &&= \prod_{k=1}^{N} \sum_{j=1}^L p_jV_j\rho V_j^\dag, \\
    &&= \prod_{k=1}^{N} \sum_{j=1}^L p_je^{-i\lambda t/N H_j}\rho e^{i\lambda t/N H_j}, \\
    &&= \prod_{k=1}^{N} \mathcal{V}_N = \prod_{k=1}^{N} \sum_{j=1}^L p_j e^{\tau\mathcal{L}_j},
\end{eqnarray}
where the last one is in liouvillian form where $\tau = \frac{\lambda t}{N}$. To express it as unitary, we can think of the expectation value of the resultant operator $\mathbb{E}[V_k] = e^{-it/N\mathbb{E}[h_k/p_k]}$ for a single mixed unitary and then for the whole process, we have $V = \mathbb{E}[V_N\dots V_1] $. 

By telescoping lemma\ref{lem2} [also seen in Lemma 3.6\cite{chen2021concentration}] we can derive(Appendix\ref{apA}) the scaling for qDrift as:
\begin{equation}
    N = \mathcal{O}(\frac{2(t \lambda )^2}{\epsilon}).
\end{equation}
\begin{lemma}{Telescoping:}\label{lem2}
    If $\mathbb{E}[V]$ and U are bounded by operator norm $\Vert \mathbb{E}[V] \Vert$, $\Vert U \Vert$ $\leq 1$, then:
    $$ \Vert \mathbb{E}[V]^N - U^N \Vert \leq N\Vert \mathbb{E}[V] - U \Vert. $$
\end{lemma}
Notably, the number of gates required for a fixed error threshold for qDrift is independent of L, which is the number of terms in the Hamiltonian. Considering an electronic system with $\mathcal{O}(N^4)$ terms, meaning $L^8$ scaling for Suzuki-trotter formulation and the fact that quantum advantage usually emerges around N $>$ 40, this randomized approach reduces the cost. We summarize this section with a flow chart in FIG.4 before moving on to random permutation.
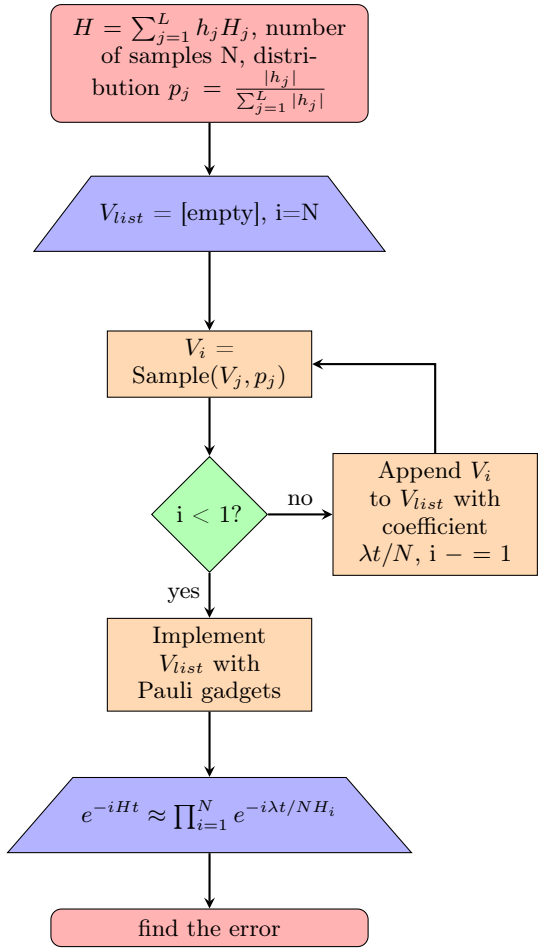
\begin{figure}
\centering
\begin{tikzpicture}[node distance=1cm]

\node (start) [startstop] {$H = \sum_{j=1}^L h_j H_j$, number of samples N, distribution $p_j = \frac{|h_j|}{\sum_{j=1}^L |h_j|}$};
\node (in1) [io, below of=start,yshift=-1cm] {$V_{list}$ = [empty], i=N};
\node (pro1) [process, below of=in1,yshift=-1cm] {$V_i$ = Sample({$V_j, p_j$})};
\node (dec1) [decision, below of=pro1, yshift=-1cm] {i < 1?};

\node (pro2a) [process, below of=dec1, yshift=-1cm] {Implement $V_{list}$ with Pauli gadgets};

\node (pro2b) [process, right of=dec1, xshift=2cm] {Append $V_i$ to $V_{list}$ with coefficient $\lambda t/N$, i $-=$ 1};
\node (out1) [io, below of=pro2a,yshift=-1cm] {$e^{-iHt} \approx \prod_{i=1}^{N}e^{-i\lambda t/N H_i}$};
\node (stop) [startstop, below of=out1,yshift=-0.5cm] {find the error};

\draw [arrow] (start) -- (in1);
\draw [arrow] (in1) -- (pro1);
\draw [arrow] (pro1) -- (dec1);
\draw [arrow] (dec1) -- node[anchor=east] {yes} (pro2a);
\draw [arrow] (dec1) -- node[anchor=south] {no} (pro2b);
\draw [arrow] (pro2b) |- (pro1);
\draw [arrow] (pro2a) -- (out1);
\draw [arrow] (out1) -- (stop);

\end{tikzpicture}
\caption{Flow chart for qDrift protocol}
\end{figure}

\subsection{Random permutation}

We first give the scaling relation for the algorithm\cite{childs2019faster}:
\begin{theorem}\label{thm2}
    Given $H = \sum_{k=1}^L h_k H_k$ as the Hamiltonian and $U = e^{-iHt}$ the evolution operator for any time t $\in$ $\mathbb{R}$. Let $S_1(t) = \prod_{k=1}^L \exp ( -i h_k H_k t)$ denotes forward lie-trotter evolution and $S_1^{rev}(t) = \prod_{k=L}^1 \exp ( -i h_k H_k t)$ represents backward evolution. We have:
    % \begin{eqnarray}
    % d_\diamondsuit (\mathcal{U},\frac{1}{2^N}(\mathcal{S}(\delta t)+\mathcal{S}^{rev}(\delta t))^N) \leq 
    % &&\frac{(\Lambda t L)^4}{N^3} \exp(\frac{2(\Lambda t L)}{N}) 
    % &&+ \frac{2(\Lambda t L)^3}{3N^2} \exp(\frac{\Lambda t L}{N})
    % \end{eqnarray}
    $$
    d_\diamondsuit (\mathcal{U},\frac{1}{2^N}(\mathcal{S}(\delta t)+\mathcal{S}^{rev}(\delta t))^N) \leq 
    \frac{(\Lambda t L)^4}{N^3} \exp(\frac{2(\Lambda t L)}{N}) $$
    $$\hspace{4.5cm}+ \frac{2(\Lambda t L)^3}{3N^2} \exp(\frac{\Lambda t L}{N}),
    $$
    where $\delta t = \frac{t}{N}$ with $N_t$ being Trotter steps and $\Lambda := \max_k \Vert h_k \Vert$.
\end{theorem}
To see it intuitively, we consider Taylor expansion for the ideal evolution of Hamiltonian $H = A+B$:
\begin{equation}
    U = e^{(A+B)t} = I + (A + B)t + \frac{1}{2} (A^2 + AB + BA + B^2)t^2 + O(t^3).
\end{equation}
Then, if we evaluate similarly the expansion for first-order lie-trotter:
\begin{equation}
    S_1(t) = e^{At}e^{Bt} = I + (A + B)t + \frac{1}{2} (A^2 + 2AB + B^2)t^2 + O(t^3).
\end{equation}
\begin{equation}
    S_1^{rev}(t) = e^{Bt}e^{At} = I + (A + B)t + \frac{1}{2} (A^2 + 2BA + B^2)t^2 + O(t^3).
\end{equation}
We can easily see the second order terms cancel, and effectively, the scheme improved the simulation to third order:
\begin{equation}
    e^{(A+B)t} = \tfrac{1}{2} e^{At}e^{Bt} + \tfrac{1}{2} e^{Bt}e^{At} + \mathcal{O}\left(t^3\right).
\end{equation}
Alternatively, equivalently, the summation forms a CPTP map:
\begin{equation}
    \rho \rightarrow \tfrac{1}{2} (e^{iBt}e^{iAt} \rho e^{-iAt} e^{-iBt}) + \tfrac{1}{2} (e^{iAt}e^{iBt} \rho e^{-iBt} e^{-iAt}).
\end{equation}
But in general, the summation of two operations is not unitary, which means we need to use mixing lemma, i.e. randomization, to sample a permutation sequence from the pool ${\pi_k} \in Sym(L)$, where index k indicates a particular sequence. After obtaining the permutation list, we then put the operator in the circuits; for example, with time interval $\tau$ and [3, 1, 2, 4, 6, 5] being chosen, we evolve with:
\begin{equation}
    V_{circuit} = e^{i\tau H_5}e^{i\tau H_6}e^{i\tau H_4}e^{i\tau H_2}e^{i\tau H_1}e^{i\tau H_3}.
\end{equation}
It is applicable for higher order Suzuki-trotter formula\cite{childs2019faster} and recently, a protocol\cite{ouyang2020compilation} called SparSto combining qDrift and random permutation has narrowed the error bound empirical by a significant margin with convex optimization in FIG.5, which leads directly to extrapolation for large system size. However, though some randomness is added in the random permutation, the dependence on $ L^2 $ still exists. On the contrary, qDrift with uniform sample distribution has escaped the problem.
\begin{figure}\label{5}
\centering
\begin{tikzpicture}[node distance=1cm]

\node (start) [startstop] {$H = \sum_{i=1}^L h_i H_i$, random permutation list $L_{perm}$ and the reverse order $L_{perm}^{rev}$, distribution $\{p_j\}$};
\node (in1) [io, below of=start,yshift=-1cm] {take $L_{perm_i}$ for i in len($L_{perm}$)};
\node (pro1) [process, below of=in1,yshift=-1cm] {x=random(0,1)};
\node (dec1) [decision, below of=pro1, yshift=-1cm] {x > $p_i$?};

\node (pro2a) [process, below of=dec1, yshift=-1cm] {Apply $H_i$ to circuit};

\node (pro2b) [process, right of=dec1, xshift=2cm] {Implement nothing new in circuit};
\node (out1) [io, below of=pro2a,yshift=-1cm] {Repeat above for the rest of the list as well as $L_{perm}^{rev}$};
\node (stop) [startstop, below of=out1,yshift=-0.5cm] {find the error};

\draw [arrow] (start) -- (in1);
\draw [arrow] (in1) -- (pro1);
\draw [arrow] (pro1) -- (dec1);
\draw [arrow] (dec1) -- node[anchor=east] {yes} (pro2a);
\draw [arrow] (dec1) -- node[anchor=south] {no} (pro2b);
\draw [arrow] (pro2b) |- (pro1);
\draw [arrow] (pro2a) -- (out1);
\draw [arrow] (out1) -- (stop);

\end{tikzpicture}
\caption{Flow chart for SparSto protocol}
\end{figure}
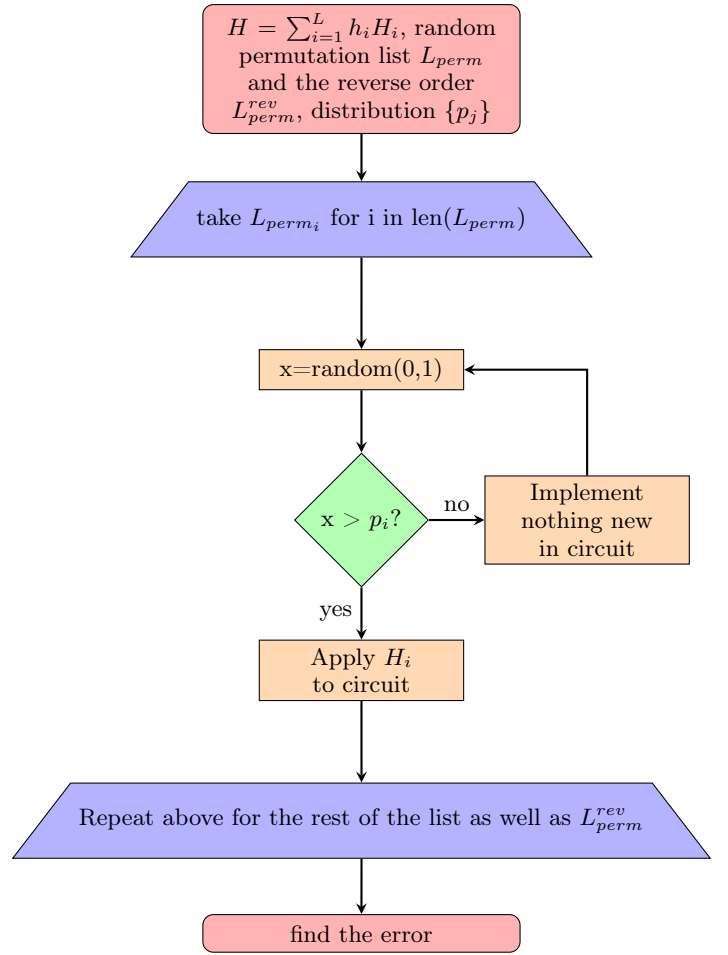

Because there is no fixed order of gates as trotterization in qDrift, we use the wording sample step instead of the time step and carefully choose the sample number to match the total time simulation, i.e., the same length of the bar. However, the individual duration of the gate might be different. With this implementation, we require fewer gates aiming for the same precision. The reason behind how randomness works better partly lies in the fact that coherent noise effects are washed out into less harmful stochastic noise\cite{wallman2016noise}.

\section{Particle number conservation} \label{sec4}

In this section we introduce our improvements over original qDrift. Our proposal aims to sample term over all Pauli stings in the physical (fermionic) space. Notably, we consider the conservation of particle number, i.e. electrons, throughout evolution. We first illustrate with examples how to keep the total number of electrons constant by applying only terms that make physical sense. Then, we will show different sampling tricks to alter the probability distribution to improve the empirical result.

\subsection{Pauli grouping}

With qDrift we sample a single Pauli string at one sample step $N_{q_i}$. For example, when we compare the spectral error with the actual time evolution
the operator at the second sample step, we might implement on the circuit Pauli strings' ZII' and 'XZX' with the first string from the part of the number counting group or the coulomb group
from TABLE.\ref{table:ta1} and the second one from part of the excitation group after the JW transformation. 
Note that the actual Pauli gates come in pairs; that is, for each hydrogen in a chain, there are electron spins up and down: above, we have three hydrogens in a chain, so we have generalized 'ZIIIII' and 'IZIIII' to just 'ZII', for instance.

It is easy to see that the \textit{Particle number operator} does not commute here with the simulated unitary using commutativity Theorem\ref{thm3}, meaning that the particle number is not conserved using Ehrenfest theorem:
\begin{equation}
    \frac{d}{dt} <\hat{P}> = \frac{1}{ih}<[\hat{P},H]>,
\end{equation}
where $\hat{P} = \sum_{i=1}^{N} (I - Z_p) $.
\begin{theorem}\label{thm3}
    For a Pauli string with $\hat{P} = \bigotimes_{i=1}^n p_i$ length n, and each $p_i \in \{X,Y,Z,I\}$, commutativity between two strings is conveniently determined by counting the number of positions in which the
    corresponding Pauli operator in subset ${X,Y,Z}$ differ. The operators commute if the total count is even; otherwise, they anticommute. 
\end{theorem}
However, we should sample each group accordingly and apply each Pauli string with the Pauli gadget. In that case, the total number of particles is constant because the individual group represents a physical process.
\begin{figure}
    \centering
    \includegraphics[scale=0.32]{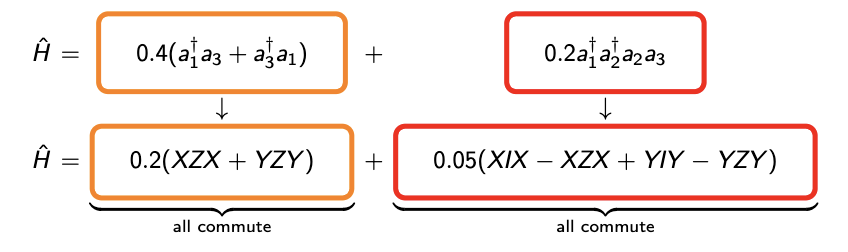}
    \caption{example of physDrift sampling scheme where Pauli strings comprising the same physical (particle number conserving) term have the same coefficient will be sampled. Given enough samples, this process will converge onto a particle conserving unitary.}
    \label{FIG.6}
\end{figure}
Another advantage of this protocol is that all Pauli strings commute inside each group as in FIG.\ref{FIG.6}. Ordering of the exponentials within each physical term will not affect the error, allowing us instead to focus on choosing an order that will optimize circuit depth, e.g., by maximizing cancellation of CNOT gates\cite{gui2021term}.
So, as long as the spectral error is taken after the whole group is applied(which can be done as there are no rules governing the exact sample steps to take error analysis), we can restrict the evolution in physical space.
Nevertheless, can we do better$?$ Because there are only five physical groups in the second quantization, which is quite small compared to the pool of qDrift. So, the strong terms with large coefficients(we give the strength of each group in FIG.7) might be averaged out by weak Pauli strings in the same group. If the probability does not differ much, this is effectively the random permutation method. So, we need to split the groups again and rearrange the strings according to each molecular orbit(or site in Heisenberg's model). This basically restricts the evolution to a physical subspace $\mathcal{H}_{phys}$:
\begin{equation}
    \mathcal{H}_{phys} = \cap_i^m \mathcal{K}er(\hat{\mathcal{P}}_i),
\end{equation}
where m is the number of molecular orbits(equals to half of the number of qubits) and the kernel space of particle number operator at each orbit: $\mathcal{K}er(\hat{\mathcal{P}_i}) = \{\ket{\psi}: (\hat{\mathcal{P}_i}-N_i) \ket{\psi} = 0\}$ where $\hat{\mathcal{P}_i}$ is the particle counting operator and $N_i$ associated value.

\begin{figure*}\label{fig7}

    \begin{tikzpicture}[node distance=0pt,
SA/.style = {single arrow, draw=blue!40!gray, very thick, fill=blue!20!gray!10,
            inner xsep=0pt, text width=\n1-4ex, align=center,% 4ex compensate vertical space above and below itemize
            font=\small
            }       
                    ]

\node (n)   [text width=0.8\textwidth, font=\large]
{
\small $$n_p = a_p^\dag a_p = \tfrac{1}{2}(\underbrace{I}_{\text{ignore}} - Z_p)$$

$$n_p n_q = \left(\underbrace{I}_{\text{ignore}}-Z_p - Z_q +Z_pZ_q \right)$$

$$a_p^\dag a_q + a_q^\dag a_p = \tfrac{1}{2}  \left(\prod_{i=p+1}^{q-1} Z_i\right) \left(X_p X_q + Y_p Y_q\right)$$

$$a_p^\dag a_q^\dag a_q a_r = \frac{1}{2}\left(\prod_{j=r+1}^{p-1} Z_j \right)\left( X_pX_r + Y_pY_r\right)\left(\frac{I-Z_q}{2}\right)$$

\begin{align*}
a^\dagger_p a^\dagger_q a_r a_s = \tfrac{1}{8} \left(\prod_{j=s+1}^{r-1} Z_j\right) \left(\prod_{k=q+1}^{p-1} Z_j\right) &\left( XXXX - XXYY + XYXY + YXXY + YXYX - YYXX + XYYX + YYYY \right)
\end{align*}
};
\path   let \p1 = ($(n.north)-(n.south)$),
            \n1 = {veclen(\y1,\x1)} in
        node[SA, rotate=90,  xshift=0.45*\n1,
             above=of n.south west] {Increasing magnitude};
        % node[SA, rotate=-90,  xshift=-0.55*\n1,
        %      above=of n.south east] {Transparency};
\end{tikzpicture}
\caption{Strength ordering of all Hamiltonian sub-groups, the bottom is the strongest.}
\end{figure*}

Given the partition scheme for the Hamiltonian, we need to find the optimized probability distribution. We have made a comparison between two proposals:
\begin{itemize}
    \item \textit{Absolute weights}: Assigning each of the Pauli strings with weight $w_i = abs(h_i)$ similarly as in qDrift. For each group $\mathcal{G}_j$ sampled, the probability is $\mathcal{A}_j = \sum_i w_i$
    \item \textit{Mean weights}: With $\mathcal{M}_j = \sum_i h_i$, we take the sign of coefficients into consideration.
\end{itemize}

Intuitively, with the \textit{Absolute weights} scheme, we are still sampling according to how strong the sub-terms are in the Hamiltonian. This means that with N being large, the histogram of Pauli string counts will converge to the one for qDrift(which will also 'drift' towards particle conserving evolution in this sense). We demonstrate this effect with a histogram plotted in FIG.8. However, adapt \textit{Mean weights}, Pauli strings with similar physical meaning but opposite signs will dilute the overall effect. For example, $h_{pqqp}$ in the Coulomb group and $h_{pqpq}$ in the spin exchange group are in the same sample space as explained above. However, the sign will be opposite due to symmetric consideration. $|h_{pqqp} + h_{pqpq}| \leq |h_{pqqp}| + |h_{pqpq}|$  with the equality achieved when $h_{pqqp} * h_{pqpq} > 0$.

\begin{figure}
    \centering
    \includegraphics[scale=0.25]{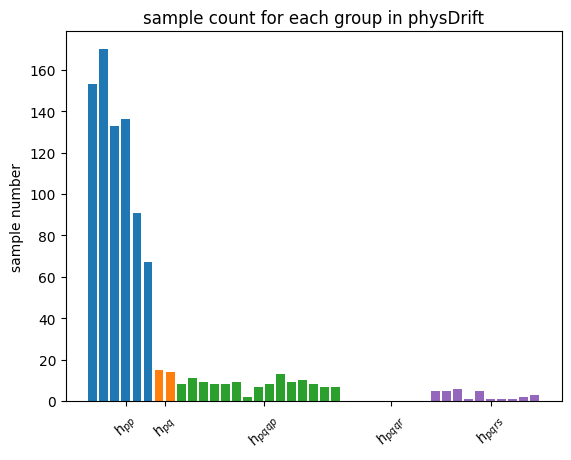}
    \includegraphics[scale=0.25]{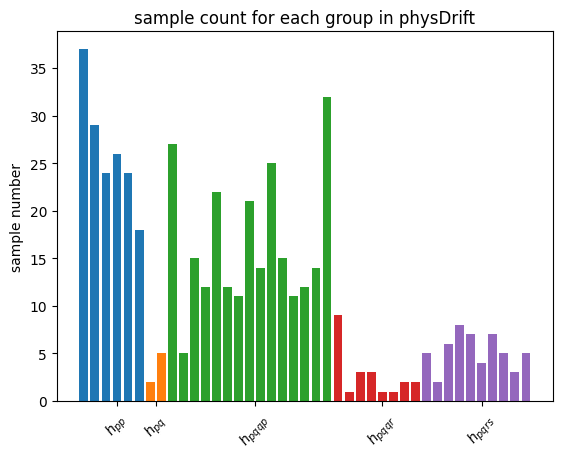}
    \includegraphics[scale=0.25]{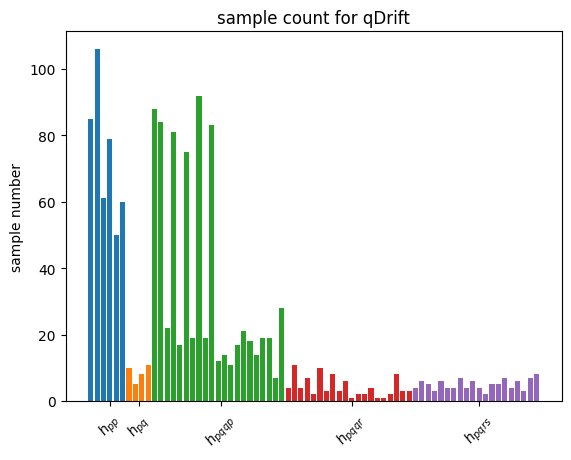}
    \caption{Histogram for a 3-hydrogen chain with 1420 sample steps, given the top left is under the mean weight scheme, and the top right is the absolute scheme. Bottom one is qDrift. For qDrift, all the Pauli strings that could be sampled are given on the x-axis; for physDrift, all the physical terms (comprising multiple Pauli strings) that could be sampled are given on the x-axis. The ordering of the terms is the same, so we can directly compare the shape of the histograms.}
    \label{FIG9}
\end{figure}

However, empirically taken \textit{Absolute weights}, the spectral error is much worse than \textit{mean weights} and worse than the original qDrift; we will use the latter for the rest of the analysis.

\subsection{Enforcing Particle Number Conservation via Lagrange Multipliers}

While physDrift naturally reduces the probability of sampling terms that do not conserve particle number, it does not explicitly enforce strict conservation. To ensure that the evolution remains within the physical subspace, we introduce a constrained optimization framework where the probability distribution $p_j$ of sampled Hamiltonian terms is modified under a conservation constraint.

\subsubsection{Constrained Sampling Formulation}

To enforce conservation, we introduce the particle number operator:
\begin{equation}
    P = \sum_{i} (I - Z_i),
\end{equation}
and require that the sampled terms satisfy:
\begin{equation}
    [H_j, P] = 0, \quad \forall j.
\end{equation}
We impose this constraint through a Lagrange multiplier $\lambda$ and solve:
\begin{equation}
    \min_{\{p_j'\}} \sum_{j} p_j' \log p_j'
\end{equation}
subject to:
\begin{align}
    \sum_{j} p_j' &= 1, \\
    \sum_{j} p_j' C_j &= 0,
\end{align}
where $C_j = \text{Tr}([H_j, P]^2)$ quantifies the degree of particle number violation. The corresponding Lagrangian is:
\begin{equation}
    \mathcal{L} = \sum_j p_j' \log p_j' + \lambda_1 \left( \sum_j p_j' - 1 \right) + \lambda_2 \sum_j p_j' C_j.
\end{equation}
Solving $\frac{\partial \mathcal{L}}{\partial p_j'} = 0$, we obtain:
\begin{equation}
    p_j' = e^{-\lambda_1 - 1 - \lambda_2 C_j}.
\end{equation}
Using normalization, $e^{\lambda_1 + 1} = \sum_j e^{-\lambda_2 C_j}$, the modified probability distribution is:
\begin{equation}
    p_j' = \frac{e^{-\lambda_2 C_j}}{\sum_k e^{-\lambda_2 C_k}}.
\end{equation}
For large $\lambda_2$, terms violating particle conservation are exponentially suppressed.

\subsection{Penalty Function Approach}
An alternative to enforcing strict conservation via constraints is to introduce a penalty term into the Hamiltonian:
\begin{equation}
    H' = H + \alpha P^2,
\end{equation}
where $\alpha$ is a tunable penalty coefficient. This modifies the coefficients in the Pauli decomposition:
\begin{equation}
    h_j' = h_j + \alpha \text{Tr}(P^2 H_j).
\end{equation}
The qDrift sampling probabilities are then updated as:
\begin{equation}
    p_j' = \frac{|h_j'|}{\sum_k |h_k'|}.
\end{equation}
This approach effectively suppresses non-conserving terms by increasing their energy cost.

\subsection{Alternative Justification: Perturbation Analysis of Unphysical Terms}
To further justify the advantage of physDrift over qDrift, we analyze the eigenvalue shifts induced by unphysical terms in the Hamiltonian. Define $H_{\text{phys}}$ as the physical Hamiltonian and $H_{\text{leak}}$ as the unphysical contribution:
\begin{equation}
    H = H_{\text{phys}} + \epsilon H_{\text{leak}}.
\end{equation}
The leading-order perturbative correction to the eigenvalues of $H$ is given by:
\begin{equation}
    \Delta E_n^{(1)} = \langle \psi_n | H_{\text{leak}} | \psi_n \rangle.
\end{equation}
Since qDrift samples all terms with probability $\mathcal{O}(\epsilon)$, the expected deviation in energy is:
\begin{equation}
    \mathbb{E}[\Delta E] = \sum_j p_j h_j.
\end{equation}
In contrast, under physDrift, $p_j'$ is exponentially suppressed for unphysical terms, yielding:
\begin{equation}
    \mathbb{E}[\Delta E]_{\text{physDrift}} \approx e^{-\lambda_2} \mathbb{E}[\Delta E]_{\text{qDrift}}.
\end{equation}
Thus, energy conservation violations are exponentially smaller under physDrift.

\subsection{Experiment results}

In this section, we will compare our algorithm with the deterministic and randomized approaches. But before showing the results, we will elaborate more on how we integrate a state-of-the-art noisy model and relevant mitigation techniques.

\subsubsection{Noise model}

We categorize the noise into three types:

\begin{enumerate}
    \item \textbf{Sampling noise}: generated by the order of unitaries applied. However, we assume that any coherent errors have been removed with the randomized compilation method, leaving only depolarising errors.
    \item \textbf{Depokarising noise}: We consider stochastic error in each gate, which can be simulated by randomly adding a Pauli operator with probability $p$ after each gate -- $p$ was chosen to reflect current ion-trap hardware.
    \item \textbf{Shot noise}: The effect of these is to effectively reduce the amplitude of the measured expectation value by an exponential factor $e^{-t}$, where $t$ is proportional to the depth of the circuit (i.e. several Pauli exponentials).
\end{enumerate}

We can then assume the simulations run with no noise. However, the shot noise $\epsilon_{\text{shot}}$ becomes correlated with the circuit depth\cite{granet2023continuous}, causing the number of shots  $N_{\text{shot}}$ to scale exponentially with the number of exponentials:
\begin{equation}
    \epsilon_{\text{shot}} \sim \frac{1}{N_{\text{shot}}^{1/2} e^{-t}} \implies N_{\text{shot}} \sim \frac{1}{\epsilon_{\text{shot}}^2 e^{-2t}}.
\end{equation}
So, we will track the spectral error for each algorithm at the same circuit depth. It is not trivial to note that we exchange circuit depth here with the number of Pauli exponentials or Pauli gadgets. This is true provided there is no optimization scheme applied\cite{Kissinger_2020, hastings2014improving}.

\subsubsection{Symmetric protection}

We restrict ourselves to the deterministic picture. By exploiting symmetries of the system, we can substantially reduce the total error $\epsilon$ of the simulation without significantly increasing the gate count\cite{Tran_2021}. Hamiltonian is invariant under the Particle number operator:
\begin{equation}\label{42}
    [H, \hat{\mathcal{P}}] = 0.
\end{equation}
Because the identity operator always commutes with other Pauli strings with the same dimension, here $\hat{\mathcal{P}}$ can be simplified to $\{ exp(i \phi Z) : \phi \in [0, 2\pi) \}$ where $\phi$ is taken randomly between [0, 2] as a multiple of $\pi$.
Consider each Trotter step $\mathcal{S}_{\delta t} = e^{-iH_{eff}t}$ where $H_{eff} = H + \delta H$ the second term being the error.
\begin{equation}
    e^{-iHt} \implies e^{-iH_{eff}t} = e^{-i( H + \delta H )t}.
\end{equation}
We then extend the original circuit with the following protection:
\begin{equation}
    V = \prod_{k=1}^{N_t} \hat{\mathcal{P}}^{\dag} \mathcal{S}_{\delta t} \hat{\mathcal{P}} = \prod_{k=1}^{N_t} e^{-i( H + \hat{\mathcal{P}}^{\dag} \delta H \hat{\mathcal{P}})t}.
\end{equation}
The second equality is from equation \ref{42}. Moreover, the error term will be reduced, similar to the average distance in a random walk. However, we argue that this improvement will not be suitable for a randomized approach because there is no fixed known unitary in the circuit, which means the error term differs upon sampling.

\subsubsection{Main result}

\begin{figure}
    \centering
    \includegraphics[scale=0.27]{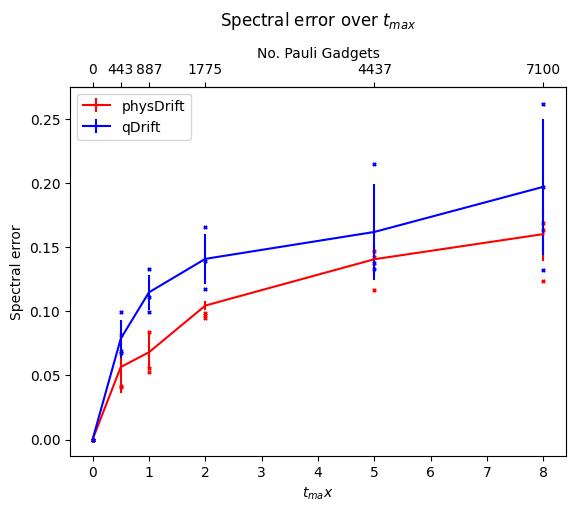}
    \includegraphics[scale=0.27]{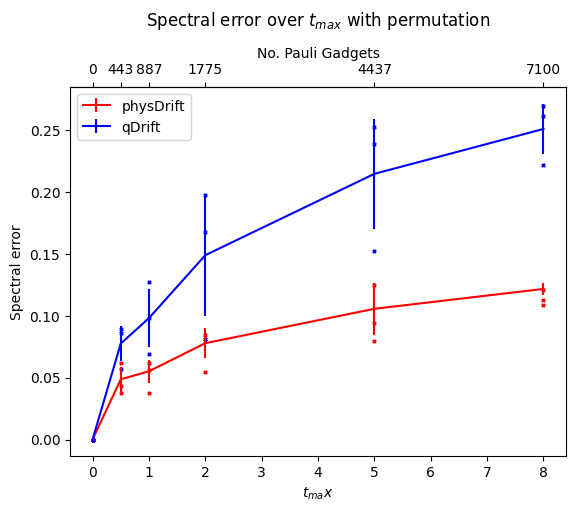}
    \caption{With $t_{max}$ taken at 0.5, 1, 2, 5 and 8, the spectral error of the average of 3 qDrift sampling sequence is evaluated on the left blue curve. The red plot indicates the averaged value for physDrift, but note that because there is no one-to-one correspondence between $t_{max}$ and sample steps, as the number of Pauli exponentials at a particular time changes for each experiment, we average the mean value of the unitary and compare it with the exact evolution. On the left, to combine physDrift with random permutation, we randomly permuted the ordering of Pauli strings. The top x-axis represents the number of Pauli exponentials in the circuit.}
    \label{dt8}
\end{figure}

Now we would like to simulate for some time $t_{max}$, and we can only use $N$ exponentials (since the noise on the hardware limits the depth), for which We assume that typical hardware can use $10^3$ or so CNOT gates. We compared the precision we can get for algorithms mentioned so far without depolarising error first in FIG.\ref{dt8}. Note that the Hamiltonian system is a 3-hydrogen chain system if not specified.

\begin{figure}
    \centering
    \includegraphics[scale=0.27]{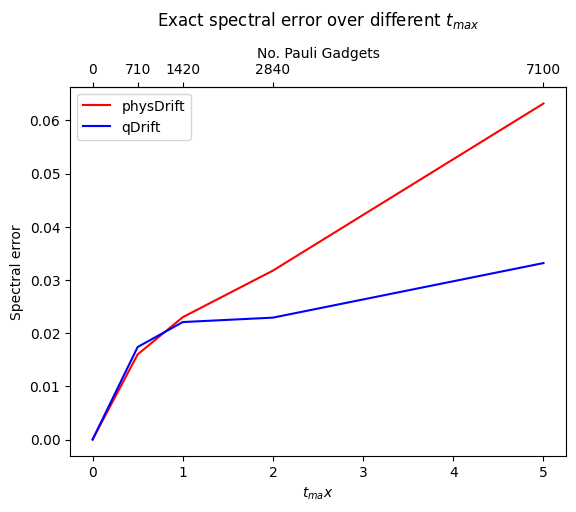}
    \includegraphics[scale=0.27]{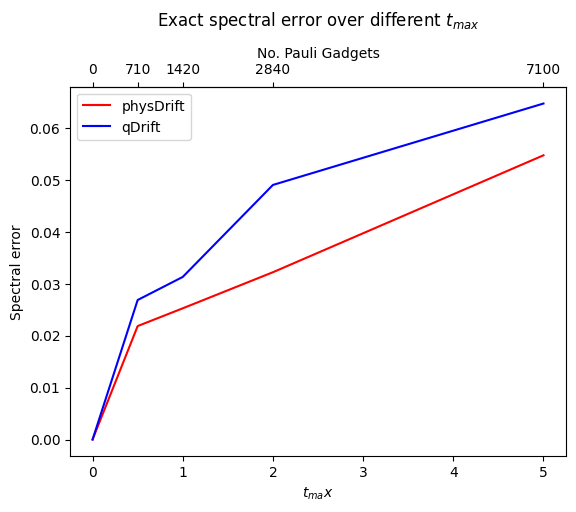}
    \caption{Now $t_{max}$ is only taken at 0.5, 1, 2, 5. The horizontal error comes from the fact that the length of the physical group is different, and because at each $t_{max}$, this only differs by a few strings, the error is too small to see. The right one is taken after applying the permutation average.}
    \label{exc3}
\end{figure}

We can see that with our scheme, the spectral error is lower and performs better when the permutation is taken. However, this has some counter effect on qDrift. To make a better analysis, we next move on comparing the exact evolution at the time where each whole group in the physDrift is applied with the simulation in FIG.\ref{exc3}

\begin{figure}
    \centering
    \includegraphics[scale=0.27]{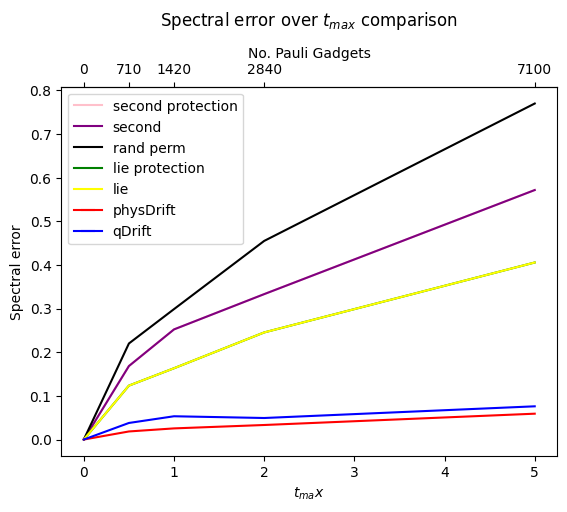}
    \includegraphics[scale=0.27]{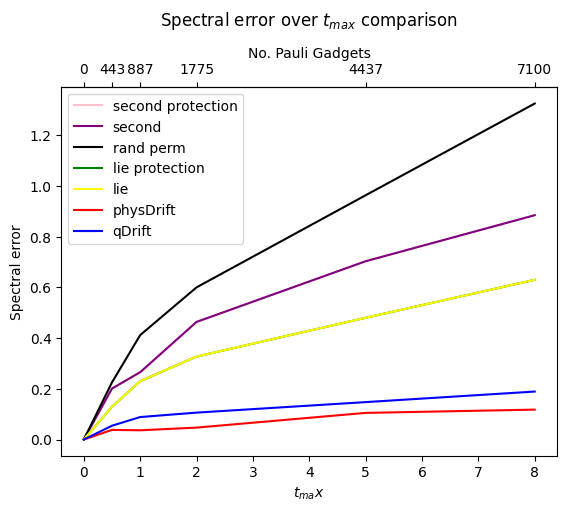}
    \caption{Each method is indicated by a different colour. The left one has a final $t_{max}$ at 5 seconds, while the right is 8 seconds. The protected case overlaps with the original algorithm without symmetric protection. The random permutation performs badly because we are only taking one single circuit with no averaging so that we can keep the depth and cost of the circuit the same.}
    \label{spec}
\end{figure}

The result for the permuted one has been improved. In contrast, the one just averaging the raw experimental result does not predict the evolution structure as accurately as qDrift. Adding in other algorithms in FIG.\ref{spec}

With the following lemma, we explain why the empirical result for physDrift is better than qDrift.

\begin{figure}
    \centering
    \includegraphics[scale=0.5]{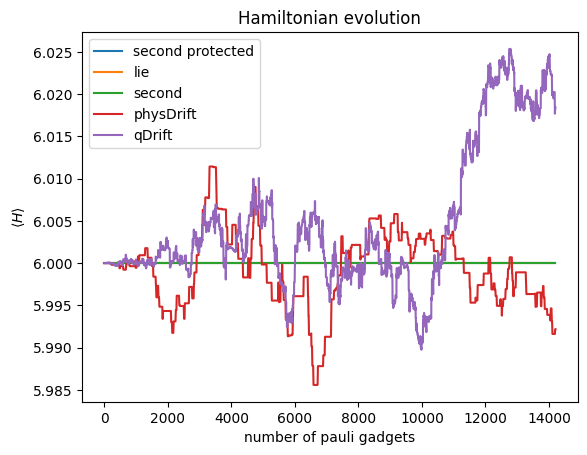}
    \caption{Evolution of the Hamiltonian. Becuase of energy conservation and $H \ket{\psi} = E \ket{\psi}$ so the expectation value should stay around E.}
    \label{H}
\end{figure}

\begin{figure}
    \centering
    \includegraphics[scale=0.5]{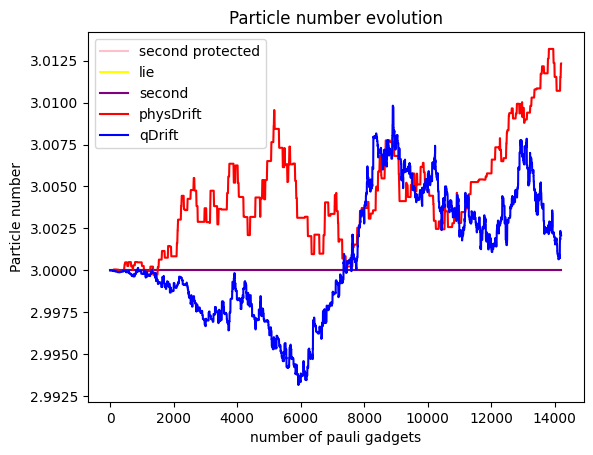}
    \caption{Particle number as time progresses. We started with three hydrogens, so there should be 3 electrons in total.}
    \label{ppp}
\end{figure}

\begin{lemma}\label{lem3}
Given a pool of $\mathcal{R}(\mathcal{L}_1,\mathcal{L}_2,\dots,\mathcal{L}_{L})$ to implement random samples of unitary, the spectral error is always greater than or equal to the protocol with a pool of $\hat{\mathcal{R}}(\mathcal{L}_1,\mathcal{L}_2,\dots,\mathcal{L}_i+\mathcal{L}_j,\dots \mathcal{L}_{\hat{L}})$ where $\hat{L} \leq L$.
\end{lemma}
\begin{proof}
    By Equation(31), $$   \Vert \mathcal{U}_N-\mathcal{V}_N \Vert_\diamondsuit = \left\Vert \sum_{n=2}^\infty \frac{t^n\mathcal{L}^n}{n!N^n}-\sum_{j}\frac{h_j}{\lambda} \sum_{n=2}^\infty \frac{\lambda^n \tau^n \mathcal{L}_j^n}{n!N^n} \right\Vert_\diamondsuit$$
    $$ \leq \sum_{n=2}^\infty \frac{t^n \left\Vert \mathcal{L}^n \right\Vert_\diamondsuit}{n!N^n}+\sum_{j}\frac{h_j}{\lambda} \sum_{n=2}^\infty \frac{ \lambda^n \tau^n \left\Vert \mathcal{L}_j^n \right\Vert_\diamondsuit}{n!N^n}, $$
    where the first order cancels with the choice of $\tau = \lambda t / N$. Looking at the second term, we know from subadditivity of diamond norm $\Vert A + B \Vert_\diamondsuit \leq \Vert A \Vert_\diamondsuit + \Vert B \Vert_\diamondsuit$ that $\Vert \mathcal{L}_1^n \Vert_\diamondsuit + \Vert \mathcal{L}_2^n \Vert_\diamondsuit + \dots \Vert (\mathcal{L}_i+\mathcal{L}_j)^n \Vert_\diamondsuit + \dots + \Vert \mathcal{L}_{\hat{L}}^n \Vert_\diamondsuit = \sum_j^{\hat{L}} \Vert \mathcal{L}_{\hat{j}}^n \Vert_\diamondsuit \leq \Vert \mathcal{L}_1^n \Vert_\diamondsuit + \Vert \mathcal{L}_2^n \Vert_\diamondsuit + \dots + \Vert \mathcal{L}_{L}^n \Vert_\diamondsuit = \sum_j^L \Vert \mathcal{L}_j^n \Vert_\diamondsuit$. This means that the theoretical bound will become tighter by combining more commuting terms.
\end{proof}
Besides particle number in FIG.\ref{ppp}, we also track the expectation of the $\langle H \rangle$ in FIG.\ref{H}, which, under exact evolution, should be conserved. So it is easy to see how the error fluctuations.

As we can see, the fluctuation of physDrift concentrates around the expected energy, while the one for qDrift overshoots after a while. This means physDrift has more tendency to stay in physical space $\mathcal{H}_{phys}$.

\begin{figure}
    \centering
    \includegraphics[scale=0.27]{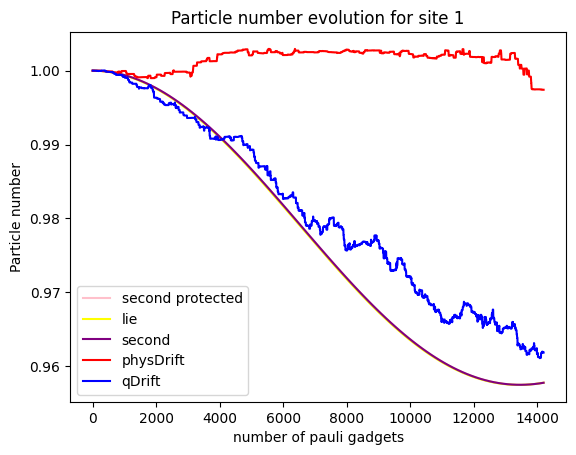}
    \includegraphics[scale=0.27]{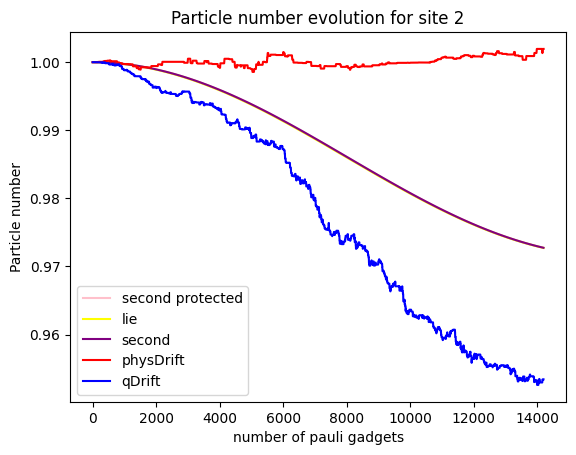}
    \includegraphics[scale=0.27]{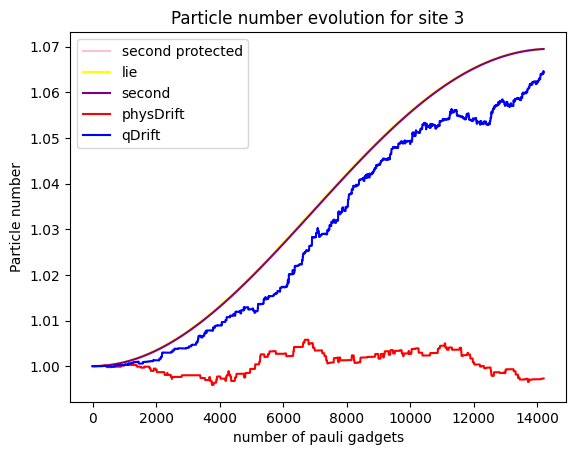}
    \caption{From left to right showing particle number evolution at orbit 1, 2 and 3}
    \label{site}
\end{figure}

To our disappointment, qDrift does slightly better again than physDrift. We have also tracked the particle number in each orbit for each algorithm shown in FIG.\ref{site}.

The simulation of an expectation value can generally be good, but the simulation of the state itself might be harmful. However, the opposite is not true.

\begin{figure}
    \centering
    \includegraphics[scale=0.5]{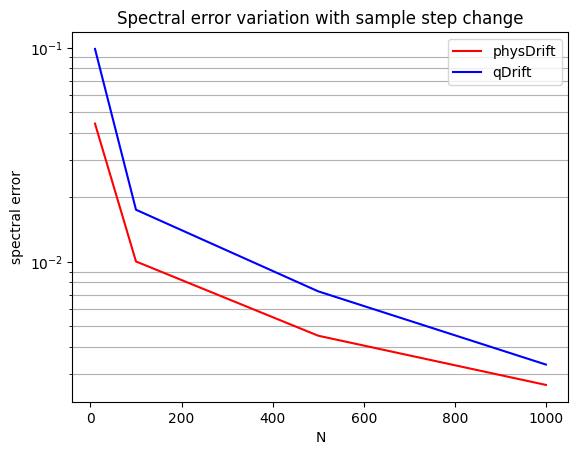}
    \caption{Plot to show in a specific case, $t_{max}=0.5s$ for a 3-hydrogen chain, how the error changes when we increase the sample steps. Note that spectral error was expressed in the log scale.}
    \label{3t05}
\end{figure}

Similar to \cite{childs2019faster}, we first tracked how the spectral error changes with the number of sample steps as in FIG.\ref{3t05}.

\begin{figure}
    \centering
    \includegraphics[scale=0.5]{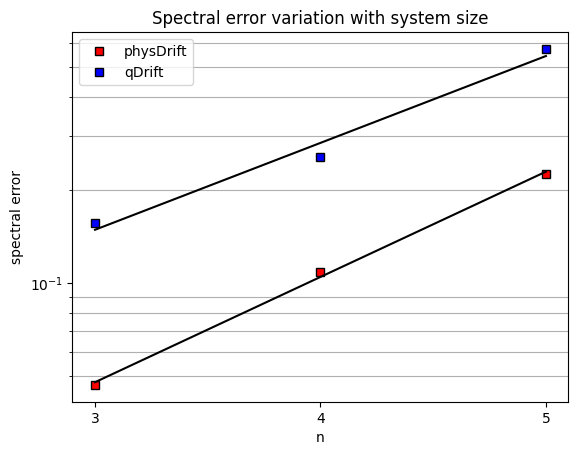}
    \caption{With a fixed number of sample steps, we extracted the error with varying system sizes. We conclude that within this scenario, physDrift generally performs better than qDrift.}
    \label{sys}
\end{figure}

Then, we plotted the system size-spectral error variation for choosing 3, 4 and 5 hydrogen chains in FIG.\ref{sys}. We should emphasize that each hydrogen has two qubits, which means the total qubits experimented with are upper bound by 12.

\begin{figure}
    \centering
    \includegraphics[scale=0.27]{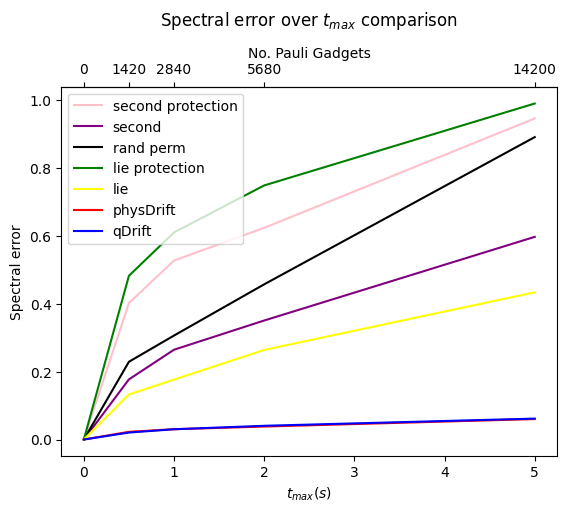}
    \includegraphics[scale=0.27]{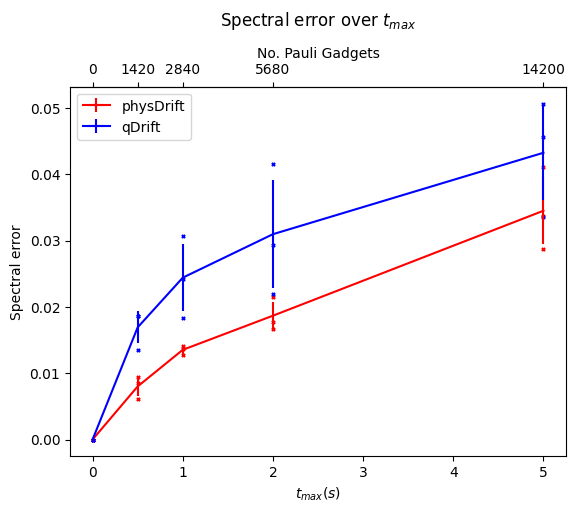}
    \includegraphics[scale=0.27]{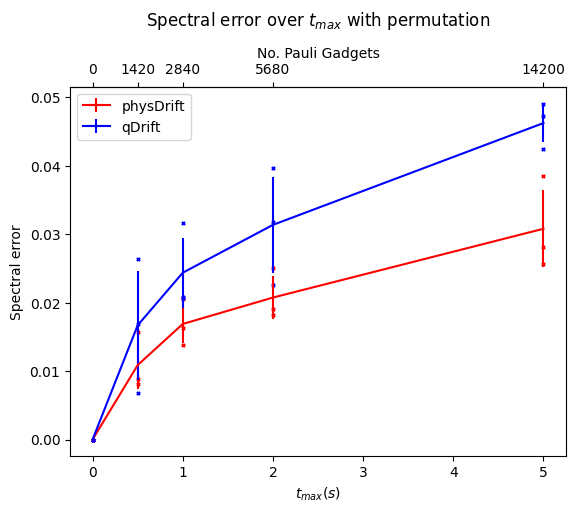}
    \caption{Top left shows the error for all algorithms while right compares physDrift and qDrift. The bottom one uses the permutation average.}
    \label{error}
\end{figure}

Now we add some depolarising error with a strength parameter around 0.1\%\cite{error}, and we can see directly that other than a minor spectral error in FIG.\ref{error}, the scheme is more tolerant to errors in the real device. This suggests that physDrfit is worth implementing on near-term quantum computers.

\section{Discussion} \label{sec5}

We show the improved quantum simulation technique with randomness based on qDrift. The basic idea is to restrict the evolution in a physical process. The result is promising because spectral error is reduced, and we have demonstrated the advantage over a naive noise model. However, we still see that sometimes, the physical property is conserved better in the existing schemes. This might result from the metrics we use not being accurate and general enough, i.e., minimizing the error or variance of expectations does not always decrease state vector error. Fundamentally, unitary evolution is the process of matrix multiplication. So exploiting mathematics behind random walk in higher dimensions, for example, with Lie group formulation, can help better understand the evolution as well\cite{versendaal2019large}. Eventually, we want to compare the result on the analogue quantum simulator and to reduce the cost transformation protocols like parity mapping\cite{seeley2012bravyi} worth investigation, which eliminates qubits due to intrinsic symmetries in the Hamiltonian.

\begin{acknowledgments}
We wish to acknowledge the support of Dr.Alex Thom and Chiara Leadbeater, who helped us discuss and suggested relevant topics.
\end{acknowledgments}

\appendix

\section{Randomized Algorithm}\label{apA}

\textbf{Lemma 1}.  
Let $\mathcal{U}=U\rho U^\dag$ and $\mathcal{V}_j=V_j\rho V_j^\dag$ be unitary channels, and let ${p_j}$ be the probability distribution for the randomized protocol, the improvements over the original mixing lemma is:\\
$$ \frac{1}{2} \Vert \mathcal{U} - \mathbb{E}[\mathcal{V}] \Vert \leq \Vert U - \mathbb{E}[V]\Vert. $$

\begin{proof}
    Let's fix a state $\ket{\psi}$ first with following notation: $\ket{u} = U\ket{\psi}$, $\ket{v} = V\ket{\psi}$. Normalization to unity ensures $|\langle u,v \rangle| \leq 1$. With Fuchs–van de Graaf relations in [\cite{FUCH},Theorem 3.33] we have:
    $$\frac{1}{2} \Vert \ket{u}\bra{u} - \ket{v}\bra{v} \Vert_1 = \sqrt{1 - \langle u,v \rangle} = \sqrt{(1-\langle u,v \rangle)(1+\langle u,v \rangle)}$$
    $$ \hspace{3cm} \leq \sqrt{2(1-Re(\langle u,v \rangle))} = \Vert \ket{u} - \ket{v} \Vert_{l_2}.$$
    In [\cite{Nielson}, Sec. 5.3], we have the fact that stabilization is not necessary for computing the diamond distance of two unitary channels:
    $$\frac{1}{2} d_\diamondsuit (\mathcal{U},\mathcal{V}) = \max_{\ket{\psi}\bra{\psi}} \frac{1}{2} \Vert (\mathcal{U}(\ket{\psi}\bra{\psi})-\mathcal{V}(\ket{\psi}\bra{\psi}) \Vert_1$$
    $$\hspace{2cm} \leq \max_{\ket{\psi}} \Vert ((U - V)\ket{\psi}) \Vert_{l_2} = \Vert U - V \Vert.$$
    However, this is only the case for a deterministic single unitary. To account for randomization with probabilistic distribution $\{p_k, V_k\}$ we use Cauchy-Schwarz:
    $$| \bra{\psi} U^{\dag} \mathbb{E}[V] \ket{\psi} |^2 = | \sum_k \sqrt{p_k}\sqrt{p_k} \bra{\psi} U^{\dag} V_k \ket{\psi} |^2,$$
    $$\hspace{2cm} = \left( \sum_k p_k \right) \sum_k p_k | \bra{\psi} U^{\dag} V_k \ket{\psi} |^2,$$
    $$\hspace{2cm} = \sum_k p_k | \bra{\psi} U^{\dag} V_k \ket{\psi} |^2.$$
    Similarly, as above, we get the following with Fuchs–van de Graaf:
    $$\frac{1}{2} \Vert \mathcal{U}(\ket{\psi}\bra{\psi})-\mathcal{V}(\ket{\psi}\bra{\psi} \Vert_1 = \sqrt{1-\sum_k p_k | \bra{\psi} U^{\dag} V_k \ket{\psi} |^2},$$
    $$\hspace{2cm} = \sqrt{1 - |\bra{\psi} U^{\dag} \mathbb{E}[V] \ket{\psi}|^2}. $$
    $$ \implies \frac{1}{2} \Vert \mathcal{U}(\ket{\psi}\bra{\psi})-\mathcal{V}(\ket{\psi}\bra{\psi} \Vert_1 \leq \Vert (U - \mathbb{E}[V]) \ket{\psi} \Vert_{l_2}.$$
    Because the average $V_k$ is not a unitary of $\mathcal{V}$ is not a CPTP map, we need a stabilization form of the diamond norm.
    $$\frac{1}{2} d_\diamondsuit (\mathcal{U},\mathcal{V}) = \max_{\ket{\psi}\bra{\psi}} \frac{1}{2} \Vert (\mathcal{U}\otimes \mathcal{I}(\ket{\psi}\bra{\psi})-\mathbb{E}[\mathcal{V}\otimes \mathcal{I}(\ket{\psi}\bra{\psi}]) \Vert_1,$$
    $$= \max_{\ket{\psi}\bra{\psi}} \frac{1}{2} \Vert (U\otimes \mathbb{I})(\ket{\psi}\bra{\psi})(U\otimes \mathbb{I})^{\dag}-\mathbb{E}[(U\otimes \mathbb{I})(\ket{\psi}\bra{\psi})(U\otimes \mathbb{I})^{\dag}] \Vert_1,$$
    $$\leq \max_{\ket{\psi}} \Vert (U\otimes  \mathbb{I} - \mathbb{E}[V\otimes \mathbb{I}])\ket{\psi} \Vert_{l_2} = \Vert (U- \mathbb{E}[V])\otimes \mathbb{I} \Vert.$$
    Finally, we can extract the identity in the product and get $\Vert (U- \mathbb{E}[V])\Vert$.
\end{proof}

\begin{center}
    \textbf{Scaling for qDrift in Sec.III.B}
\end{center}

\begin{theorem}[Taylor expansion bound]
    The error of a function \textit{f} with Taylor expansion approximation to order \textit{k} can be considered as the remainder term in order 
    $\textit{k}+1$\cite{Childs_2018}.
    $$\mathcal{R}_k(e^\alpha) \vert \leq \frac{\vert \alpha \vert^{k+1}}{(k+1)!}  e^{ \vert \alpha \vert }, \forall \alpha \in \mathbb{C},$$
    where $\mathcal{R}_k(f)$ is the remainder Taylor expansion to order \textit{k} of the function \textit{f}, for instance, $\mathcal{R}_1 (e^x)=\mathcal{R}_1 (\sum_{n=0}^\infty \frac{x^n}{n!})=\sum_{n=2}^\infty \frac{x^n}{n!}.$
\end{theorem}

We write the ideal channel as $\mathcal{U}_N = e^{\frac{t}{N}\mathcal{L}}$ and the approximation as in equation 31  $\mathcal{V}_N = \sum_{j=1}^L p_j e^{\tau \mathcal{L}_j}$ for a single sample step where $p_j = \frac{h_j}{\sum_{j=1}^L h_j}.$

Expand both channels:
$$e^{t\mathcal{L}} \approx \mathbb{I} + \frac{t}{N}\mathcal{L} + \frac{1}{2!} \frac{t^2\mathcal{L}^2}{N^2} + \dots,$$
$$\sum_{j=1}^L p_j e^{\tau \mathcal{L}_j} \approx \mathbb{I} + \sum_{j=1}^{L} p_j \tau \mathcal{L}_j + \frac{1}{2!} \sum_{j=1}^L p_j \tau^2\mathcal{L}_j^2 + \dots.$$
Now, because we can always choose $\tau = \Lambda t /N$, the first order gets cancelled exactly.
$$\Vert \mathcal{U}_N-\mathcal{V}_N \Vert_\diamondsuit = \Vert \sum_{n=2}^\infty \frac{t^n \mathcal{L}^n}{n!N^n}-\sum_{j}\frac{h_j}{\lambda} \sum_{n=2}^\infty \frac{\lambda^n t^n \mathcal{L}_j^n}{n!N^n} \Vert_\diamondsuit,$$
$$\leq \sum_{n=2}^\infty \frac{t^n\Vert \mathcal{L}^n \Vert_\Diamond }{n!N^n} + \sum_{j}\frac{h_j}{\lambda} \sum_{n=2}^\infty \frac{\lambda^n t^n \Vert\mathcal{L}_j^n \Vert_\diamondsuit }{n!N^n},$$
$$\leq \sum_{n=2}^\infty \frac{1}{n!}\left( \frac{2\lambda t}{N}\right)^n+\sum_{j}\frac{h_j}{\lambda} \sum_{n=2}^\infty \frac{1}{n!}\left( \frac{2\lambda t}{N}\right)^n,$$
$$=2\sum_{n=2}^\infty \frac{1}{n!}\left( \frac{2\lambda t}{N}\right)^n.$$
where in the third line, we used the fact that each $H_j$ is unitary to get the bound that $\Vert \mathcal{L}_ j \Vert_\diamondsuit \leq 2 \Vert H_ j \Vert \leq 2$. Similarly we have inequality $\Vert \mathcal{L} \Vert_\diamondsuit \leq 2\Vert H\Vert \leq 2\lambda$.

Apply Theorem 4 with \textit{k} = 1 and $x = \frac{2\lambda t}{N}$ now:
$$d_\diamondsuit (\mathcal{U}_N,\mathcal{V}_N) \leq \frac{2\lambda^2 t^2}{N^2} e^{2\lambda t/N}.$$
when $N >> \lambda t$, the exponential term drops out, and with the help of a telescoping lemma, we get:
$$d_\diamondsuit (\mathcal{U}_N,\mathcal{V}_N) \leq \frac{2\lambda^2 t^2}{N}.$$

\section{Optimize Trotter order}\label{apB}

From equation 7 we have $N = \mathcal{O}(\frac{(t L \Lambda )^{1+\frac{1}{2k}}}{\epsilon^{1/2k}})$ for higher order trotter scaling. To find the trade between accuracy in terms of order and the number of gates we need, we consider T the number of total gates required to have the scaling $T \propto \frac{5^k}{\epsilon^{1/2k}} $ by including the recursive formula(five terms in total) and ignoring other effects like total time and number of terms in the system Hamiltonian.

It is straightforward to calculate the first derivative of T to optimize the cost:
$$\frac{d T(2k)}{dk} \approx 2k^2log(5) - log(\frac{1}{\epsilon}).$$
For example the numerical solution with $\epsilon \approx 10^{-6}$ is around $k = 2.02$.

\section{Important Sampling}\label{apC}

Important sampling is a powerful technique to compute expectations:
$$\mathbb{E}_p[f(x)] = \sum_x p(x)f(x)$$
If we want to reduce the variance, we can re-weight accordingly:
$$\mathbb{E}_p[f(x)] = \sum_x q(x) \frac{p(x)}{q(x)} f(x) = \mathbb{E}_p[w(x)f(x)],$$
by a simple weighting scheme:
$$q_c(j) = \frac{h_j}{C_j \lambda_c}, \lambda_c = \sum_l \frac{h_l}{C_l}.$$
We can prove that the sampling requires less total simulation cost than the original one with Jensen's inequality:

$$N_{q_c} \mathbb{E}_{q_c}[C] \leq N_{p_c} \mathbb{E}_{p_c}[C].$$ for any given precision $\epsilon$, where $N_{q_c}$ is the sample steps for the re-weighted important sampling and $N_{p_c}$ is the original one. We first show the error bound for $q_c$ by expressing Hamiltonian as $H = \sum_j h_jH_j = \lambda \mathbb{E}_p[H_j] = \lambda \mathbb{E}_{q_c}[w(q)H_j]$ where $w(q) = \frac{h_j}{\lambda q(j)}$. Then:
$$U = e^{-itH} = e^{-it\lambda\mathbb{E}_p[H_j]} = ^{-it\lambda\mathbb{E}_q[w(j)H_j]} = e^{-i\mathbb{E}_q[X_j]}. $$
Note that we can obtain the bound:
$$\Vert X_j \Vert  = \frac{h_jt}{q(j)} \Vert H_j \Vert,$$

$$d_\diamondsuit (\mathcal{U}_N,\mathbb{E}_q[\mathcal{V}_N]) \leq 2\Vert U_N - \mathbb{E}_q[V_N] \Vert,$$
$$ \hspace{2cm} = 2\Vert e^{-i\mathbb{E}_q[X(t)]} - \mathbb{I} + i\mathbb{E}_q[X(t)] + \mathbb{E}_q[\mathbb{I} - iX - e^{-iX(t)}] \Vert,$$
$$ \hspace{2cm} \leq 2\Vert e^{-i\mathbb{E}_q[X(t)]} - \mathbb{I} + i\mathbb{E}_q[X(t)]\Vert + 2\mathbb{E}_q[\Vert\mathbb{I} - iX - e^{-iX(t)}\Vert],$$
$$\hspace{2cm} \leq \Vert \mathbb{E}_q[X] \Vert^2 + \mathbb{E}_q[\Vert X \Vert^2],$$
$$\hspace{2cm} \leq (t\lambda)^2 + \mathbb{E}_q[(\frac{h_jt}{q(j)})^2],$$
$$\hspace{2cm} \leq (t\lambda)^2(1 + \mathbb{E}_p[(w(j)]),$$
where $V = X(t)$ represents the mixed operator, we further relax the bound using triangular inequality in line three.
Again, using the telescoping lemma:
$$d_\diamondsuit (\mathcal{U},\mathbb{E}_q[\mathcal{V}]) \leq \frac{(t\lambda)^2}{N}(1 + \mathbb{E}_p[(w(j)]).$$

The total cost of the important distribution is:
$$C_{q_c} = N_{q_c} \mathbb{E}_{q_c}[C] = \frac{(t\lambda)^2}{\epsilon}(1 + \mathbb{E}_p[(w(j)])\mathbb{E}_{q_c}[C],$$
$$\hspace{2cm} = \frac{(t\lambda)^2}{\epsilon}(1 + \sum_j\frac{h_j}{\lambda}w(j))\mathbb{E}_{q_c}[C],$$

$$\hspace{2cm} = \frac{(t\lambda)^2}{\epsilon}(1 + \sum_j\lambda_c\frac{h_j}{\lambda^2}C_j)\mathbb{E}_{q_c}[C],$$

$$\hspace{2cm} = \frac{(t\lambda)^2}{\epsilon}(1 + \frac{\lambda_c}{\lambda}\mathbb{E}_p[C])\mathbb{E}_{q_c}[C],$$

$$\hspace{2cm} = \frac{(t\lambda)^2}{\epsilon}(1 + \mathbb{E}_p[\frac{1}{C}]\mathbb{E}_p[C])\mathbb{E}_{q_c}[C],$$

$$\hspace{2cm} = \frac{(t\lambda)^2}{\epsilon}\frac{1 + \mathbb{E}_p[\frac{1}{C}]\mathbb{E}_p[C]}{\mathbb{E}_{p}[1/C]}.$$

Comparing this to the one for qDrift $C_p = \frac{2(t\lambda)^2}{\epsilon}\mathbb{E}_p[C]$ we need:

$$\frac{1 + \mathbb{E}_p[1/C]\mathbb{E}_p[C]}{\mathbb{E}_{p}[1/C]} \leq 2\mathbb{E}_p[C],$$

$$\implies \mathbb{E}_p[1/C]\mathbb{E}_p[C] \leq 1.$$
Which is always satified by Jensen's inequality.
 
\nocite{*}

\bibliography{apssamp}% Produces the bibliography via BibTeX.

\end{document}